# Multi-Armed Bandit Learning for Content Provisioning in Network of UAVs

Amit Kumar Bhuyan, Hrishikesh Dutta, and Subir Biswas
*Electrical and Computer Engineering, Michigan State University, East Lansing, MI, USA*
*bhuyanam@msu.edu, duttahr1@msu.edu, sbiswas@msu.edu*

***Abstract* — This paper proposes an unmanned aerial vehicle (UAV) aided content management system in communication-challenged disaster scenarios. Without cellular infrastructure in such scenarios, community of stranded users can be provided access to situation-critical contents using a hybrid network of static and traveling UAVs. A set of relatively static anchor UAVs can download content from central servers and provide content access to its local users. A set of ferrying UAVs with wider mobility can provision content to users by shuffling them across different anchor UAVs while visiting different communities of users. The objective is to design a content dissemination system that *on-the-fly* learns content caching policies for maximizing content availability to the stranded users. This paper proposes a decentralized *Top-k* Multi-Armed Bandit Learning model for UAV-caching decision-making that takes geo-temporal differences in content popularity and heterogeneity in content demands into consideration. The proposed paradigm is able to combine the expected reward maximization attribute and a proposed multi-dimensional reward structure of *Top-k* Multi-Armed Bandit, for caching decision at the UAVs. This study is done for different user-specified tolerable access delay, heterogeneous popularity distributions, and inter-community geographical characteristics. Functional verification and performance evaluation of the proposed caching framework is done for a wide range of network size, UAV distribution, and content popularity.**

## I. Introduction

Disasters such as earthquakes, floods, wars, and other catastrophic events can have devastating effects on people's lives and properties, as well as communication infrastructures. In such situations, people may be forced to migrate to areas without proper communication infrastructure, leaving them without access to important information such as the state of the disaster, rescue and relief operations, weather reports, rehabilitation efforts, etc. This paper proposes the use of Unmanned Aerial Vehicles (UAVs) as an alternative content provisioning platform when fixed communication infrastructure such as cellular phone towers is unavailable. UAVs, however, bring their own limitations in storage capacity, flight time, etc., which add new challenges to UAV based content storage and dissemination system.

The paper presents a UAV-aided content caching system that uses Multi-armed Bandit Learning to perform optimal caching in communication-challenged environments. The system employs a multi-dimensional reward structure to encapsulate the holistic content request experience of a local learning model within a UAV, with the aim of maximizing reward [1] and improving content dissemination performance across UAVs. The framework focuses on scenarios where disaster/war-stricken populations are stranded and geographically clustered into multiple communities that may not have access to surviving cellular base stations. In such scenarios, the request patterns at different communities and the tolerable access delay (*TAD*) [2] can be different for different contents based on the type and urgency of the requested information. The proposed MAB learning solution deploys UAV-based tactical content service provisioning that can make caching decisions on the fly without prior knowledge of content request pattern.

The proposed content provisioning system uses a two-tier architecture consisting of anchor-UAVs (A-UAVs) and ferrying-UAVs (F-UAVs). Each disaster-isolated user community is served by an A-UAV with expensive vertical connectivity, such as a satellite link, while F-UAVs ferry and distribute content across the A-UAVs. The goal is to provide high-availability content access to all the communities without incurring the cost of excessive vertical link usage by the A-UAVs. To achieve this, the paper attempts to answer several questions, including optimal content caching policies for both A-UAVs and F-UAVs, and which content should be transferred from A-UAVs to F-UAVs. The questions are addressed with the goal of learning optimal caching policies on the fly and maximizing content availability across all user communities. The learning is explored using a novel application of *Top-k* Multi-Armed Bandit framework.

Existing work [2-5] on content caching for UAVs often considers global popularity of contents without taking geo-temporal differences (changing with time and location) in content popularity and heterogeneity in demand into account. The approaches in [6,7] employ function approximation to estimate long-term demand heterogeneities, which are sluggish in learning. This paper addresses these shortcomings by using a Multi-Armed Bandit based learning model that makes learning fast and adaptive to demand heterogeneity via information sharing among UAVs. Most works [2-7] have proposed the use of UAVs to form Ad Hoc networks that can fill communication gaps caused by infrastructure destruction, but these approaches may not work well when all communication infrastructures are destroyed, and a fully functional alternative needs to be formed. The proposed system sets out to address these shortcomings by using a *Top-k* Multi-armed Bandit Learning strategy for caching decisions at UAVs. The strategy can handle heterogeneous user demand patterns and maximize content availability to the requesting users.

The key contributions are as follows. First, a UAV-aided content caching and dissemination framework is proposed which can learn optional caching policies on the fly using a *Top-k* Multi-armed Bandit Learning. Second, a multi-dimensional reward structure for the *Top-k* MAB model is proposed based on shared information between the UAVs to improve content availability. These rewards take local and global context of content popularities into consideration while learning optimal caching policies. Third, the interactions between learnt caching policies and QoS expectation, namely, Tolerable Access Delay, is studied and characterized. Fourth, the relation between user demands and learnt caching policies are explored for learning model parameter turning. Finally, simulation experiments and analytical models are developed for functional verification and performance evaluation of the proposed caching and content dissemination framework.

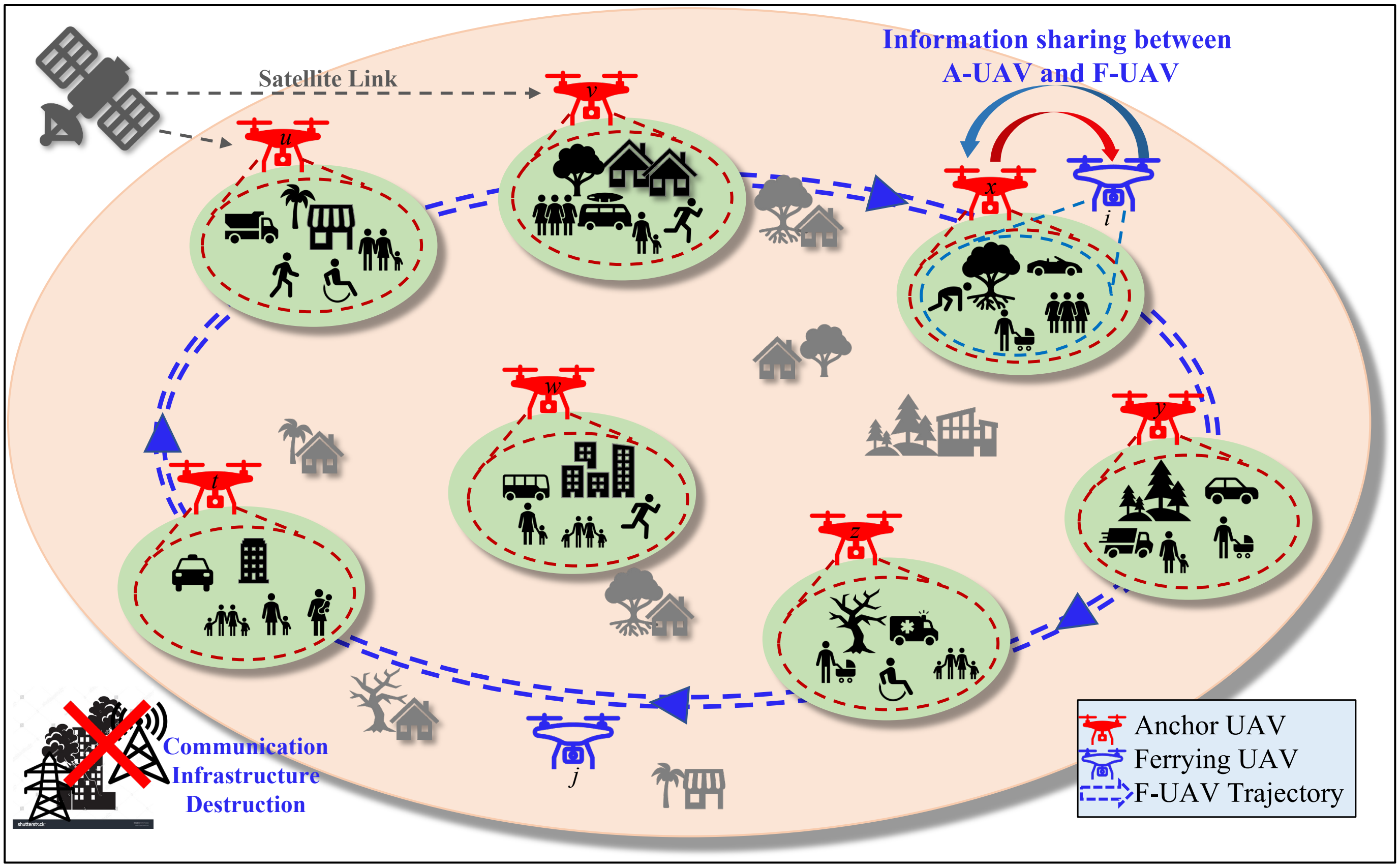

Fig. 1. Coordinated UAV system for content caching and distribution in environments without communication infrastructure

## II. System Model

### A. *UAV Hierarchy*

As shown in Fig. 1, a two-tiered UAV-assisted content dissemination system is deployed. Each community is served by a dedicated A-UAV that uses a lateral wireless connection (i.e., WiFi etc.) to communicate with users in that community. The system introduces a set of ferrying UAVs (F-UAVs), which are mobile and only have lateral communication links such as Wi-Fi. Unlike the A-UAVs, the F-UAVs do not possess vertical links. The F-UAVs act as content transfer agents across different user communities by selectively transferring content across the A-UAVs via their lateral links.

### B. *Content Demand and Provisioning Model*

The generation of content requests, content popularity distribution and quality of services are outlined below.

Content Popularity: Research has shown that content request patterns from a population often follows a Zipf distribution [4, 5], where the popularity of a content is proportional to the inverse of its rank and is a geometric multiple of the next popular content. Popularity of content '$i$' is given as:

$$p_\alpha(i) = \frac{\left(\frac{1}{i}\right)^\alpha}{\sum_{k\in C}\left(\frac{1}{k}\right)^\alpha} \quad (1)$$

The Zipf parameter, $\alpha$, determines the distribution's skewness, while the total number of contents in the pool is represented by the parameter $C$. The inter-request time from a user follows the popular exponential distribution [10].

Tolerable Access Delay: For each requested content, the user specifies a Tolerable Access Delay ($TAD$) [2], which serves as a quality-of-service parameter and represents the amount of time the user can wait before the content is downloaded.

Content Provisioning: Upon receiving a request from one of its community users, the relevant A-UAV first searches its local storage for the content. If the content is not found, the A-UAV waits for a potential future delivery by a traveling F-UAV. If no F-UAV arrives with the requested content within the specified *TAD*, the A-UAV then proceeds to download it through its vertical link, which is usually expensive.

## III. Caching Based on Content Pre-loading at Anchor UAVs

This section discusses caching policies based on content pre-loading at A-UAVs that assumes pre-assigned, static, and globally known content popularities. After understanding the limitations of these caching policies, this paper proposes a runtime, dynamic and adaptive *Top-k* Multi-armed Bandit based caching mechanism, which is explained in a later section.

### A. *Pre-loading Policies at Anchor UAVs (A-UAVs)*

The *Fully Duplicated* (*FD*) mechanism [8] is a naive approach that allows A-UAVs to download content from vertical links upon request by local users. However, the FD mechanism has limitations such as content duplication, high vertical link download costs, and suboptimal utilization of UAV cache space. *Smart Exclusive Caching* (*SEC*) [8] overcomes the limitations of the FD mechanism by storing a set number of unique contents in all A-UAVs and sharing them among communities via F-UAVs. Assuming globally known *homogeneous content popularity* across all user communities, the SEC mechanism divides the cache into two segments. Segment-1 contains the top $\lambda . C_A$ popular contents cached in all A-UAVs, while Segment-2 contains unique contents $(1 - \lambda). C_A$, where $\lambda$ is the *Storage Segmentation Factor*. Total contents in the system as per SEC is given as:

$$C_{sys} = \lambda . C_A + N_A . (1 - \lambda). C_A \quad (2)$$

*Popularity-Based Caching* (*PBC*) [10] is employed when different communities have different content preferences. PBC divides the cache space of a A-UAV into two segments, considering the *heterogeneous popularity* sequence of the local community. Segment-1 caches the most popular contents, which can be exclusive to a A-UAV ($C_E$) or non-exclusive i.e., may be cached across multiple A-UAVs ($C_{NE}$), while Segment-2 is the same as SEC. Therefore, by modifying Eq. 2, total number of contents in the system can be expressed as:

$$C_{sys} = C_{NE} + C_E^{total} + N_A . (1 - \lambda). C_A \geq \lambda . C_A + N_A . (1 - \lambda). C_A \quad (3)$$

*Value-Based Caching* (*VBC*) [10] further enhances the caching policy by storing top-valued contents in Segment-1 of A-UAV, where *value* of contents comprises of their popularity and tolerable access delay. Value of a content '$i$' be calculated as:

$$V(i) = \kappa \upsilon \times \frac{p_\alpha(i)}{TAD(i)} = \kappa \times \frac{TAD_{min}}{p_\alpha(1)} \times \frac{p_\alpha(i)}{TAD(i)} \quad (4)$$

In this equation, $p_\alpha(i)$ represents the content's popularity as per the Zipf distribution, $TAD(i)$ is the content's tolerable access delay, $\kappa$ is a scalar weight that increases as popularity decreases, and $\upsilon$ is a normalization constant. The normalization constant is calculated for a given Zipf (popularity) parameter $\alpha$ using the minimum possible $TAD$ ($TAD_{min}$ ) and the maximum possible popularity, which is $p_\alpha(1)$. The value of $V(i)$ is bounded between [0,1] and increases as $p_\alpha(i)$ increases and $TAD(i)$ decreases and can present a holistic quantifiable measure for caching decision.

The caching policy for F-UAVs remains the same for all the discussed and forthcoming caching policies for A-UAVs [8-10]. An F-UAV ferries content from already visited A-UAVs to future visiting A-UAVs in its trajectory. The caching policy of an A-UAV determines the utility of an F-UAV where every A-UAV should maintain sufficient contents in its cache space to optimize the F-UAV cache utilization.

### *B. Limitations of Cache Pre-loading at A-UAVs*

The caching policies discussed in this section rely on pre-loading content into A-UAVs, which has certain limitations. This approach assumes a priori knowledge of the popularity distribution of all the content in the system, which can hinder practical feasibility during deployment. Local popularity estimation of requested content within individual A-UAVs can partially alleviate this issue, but it cannot adjust the crucial storage segmentation factor ($\lambda$) (see section IIIA) for maximizing availability across the entire system of A-UAVs and their communities. Collaborative global popularity estimation can be introduced, but it fails to capture demand heterogeneity across different A-UAV communities.

The limitations listed above can be addressed by employing a *Top-k* Multi-armed Bandit (*Top-k* MAB) learning-based caching mechanism at the A-UAVs, which is explained in the following section. This paradigm is able to leverage the expected reward maximization attribute of MAB and intelligence sharing nature of proposed multi-dimensional reward structure for caching decision at the A-UAVs.

## IV. Decentralized Caching with Multi-Armed Bandit

Once a A-UAV is deployed into a community, its subsequent action is to decide which contents to download (via its vertical link) and cache such that content availability to the requesting users can be maximized. This goal is achieved by employing a *Top-k Multi-Armed Bandit* learning agent in the A-UAV.

### *A. Top-k Multi-Armed Bandit Learning*

Multi-Armed Bandit is a classic problem in reinforcement learning [1] and decision-making. At each round $t$, an agent chooses an arm $A_t$ out of $N$ arms, denoted by $A_1, A_2, \ldots, A_N$, and observes a reward $R_t$. Each arm $i$ has an unknown reward distribution with mean $\mu_i$ and variance $\sigma_i^2$. The agent's goal is to maximize the total expected reward $R_T$ over $T$ rounds, where $T$ is the total number of rounds (time horizon):

$$R_T = max \sum_{t=1}^{T} E[R_t] \quad (5)$$

This paper uses a variant of MAB called *Top-k Multi-Armed Bandit* [1, 11]. Here, the agent has to choose $k$ arms out of a larger set of $N$ arms, as opposed to choosing one arm in classical MAB, and receives a reward for each arm in the chosen set. The goal of the agent is to maximize the total cumulative reward $R_T$ obtained over a finite time horizon $T$:

$$R_T = max \sum_{t=1}^{T} \sum_{i=1}^{k} E[R_{i,t}] \quad (6)$$

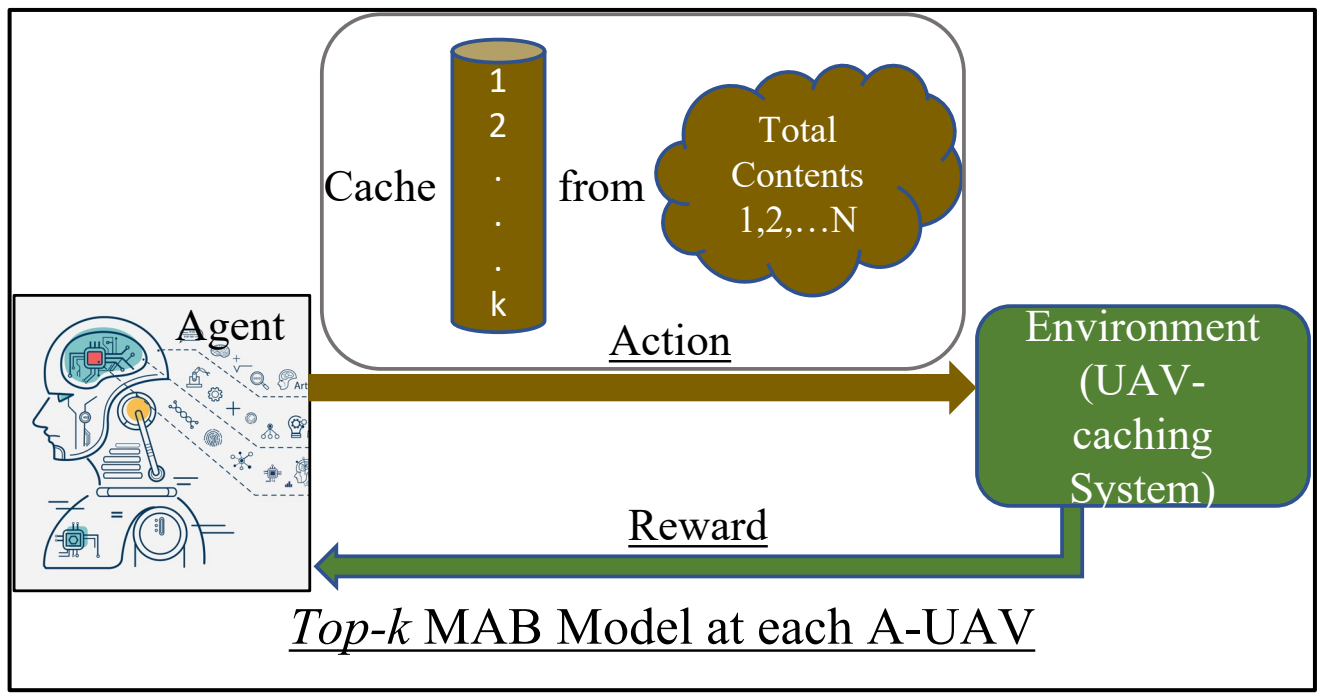

Fig. 2. *Top-k* Multi-Armed Bandit Learning for Caching Policy at A-UAVs

### *B. Decentralized Caching using Top-k Multi-Armed Bandit*

In the scenario of UAV-caching, there is a *Top-k* MAB agent in each A-UAV. Here, choosing each content for caching corresponds to choosing an arm. The '$k$' of *Top-k* MAB agent corresponds to the caching capacity of A-UAV, i.e., $k = C_A$. The agent's aim is to select '$C_A$' contents out of a larger set of '$N$' contents to be cached in an A-UAV such that content availability to the users can be maximized. Here, the UAV-aided content dissemination system is the learning environment where the A-UAVs interact through their actions of choosing specific sets of contents to be cached. The feedback from the environment for the taken actions are in the form of rewards/penalties. Actions are rewarded when cached contents are requested by the users and are served to the users within the given tolerable access delay or penalized otherwise. The top $C_A$ contents that accumulate most reward from the corresponding community and other communities are chosen to be cached at a A-UAV. It should be noted that the *Top-k* MAB agents in the A-UAVs are provided with no *a priori* information about the content popularity at the corresponding user communities.

A learning decision epoch for each *Top-k* MAB agent is set according to the F-UAVs accessibility at the corresponding community (i.e., an F-UAV's visiting frequency). This is because the F-UAVs carries the content availability information from the communities in its trajectory that is leveraged for learning at the A-UAVs' *Top-k* MAB agents using appropriately designed multi-dimensional rewards. The agent learns to cache contents via the multi-dimensional reward structure which has three parts: namely *local*, *ferrying,* and *global* reward. The first corresponds to the increase in availability at an A-UAV's corresponding community i.e., *increase in local availability* ($\delta_l$). The second is related to the contents that are cached in an A-UAV, and are responsible for increase in availability at other communities i.e., *ferried content availability* ($\delta_f$). A global reward is received when cached contents add to increase in average availability across all communities. This is called *increase in global availability* ($\delta_g$). The three types of rewards are given below:

$$R_i^L = \begin{cases} 1, & for\ \delta_l > 0 \\ -1, & for\ \delta_l < 0 \end{cases} \quad (7)$$

$$R_i^F = \begin{cases} 1, & for\ \delta_f > 0 \\ -1, & for\ \delta_f < 0 \end{cases} \quad (8)$$

$$R_i^G = \begin{cases} 1, & for\ \delta_g > 0 \\ -1, & for\ \delta_g < 0 \end{cases} \quad (9)$$

In the above equations, $R_i^L$, $R_i^F$, and $R_i^G$ are rewards according to increase in availability for content '$i$' cached in an A-UAV.

Learning is achieved using a tabular method where a Q-table is maintained for all contents in the system. The value corresponding to each content is called a Q-value or action-value [1]. The agent updates the Q-value for a content at every learning epoch according to the multi-dimensional rewards in Eqns. 7-9 from the interaction with the environment (UAV-aided content dissemination system) and learns the best actions (contents cached). The recursive expression which explains Q-value update for a content '$i$' is given as follows:

$$Q(i) \leftarrow Q(i) + \alpha\big(r(i) - Q(i)\big) \quad (10)$$

Here, $Q(i)$ represents the Q-value of a content '$i$'; $r(i)$ is the reward received by caching content '$i$'; $\alpha$ is a hyper-parameter which controls the learning rate. The Q-values for all contents are initialized with zero to ensure no *a priori* information for a *Top-k* MAB agent. Also, it ensures equal importance to all contents for caching decisions. An epsilon-greedy ($\epsilon$-greedy) exploration strategy is implemented. Such exploration strategy guarantees that every content gets to be cached in an A-UAV. As learning progresses, exploration decays and best contents with highest Q-values are exploited with the aim of maximizing accumulated reward which improves the caching policy and thus increases content availability.

The proposed algorithm enables *Top-k* MAB agents in A-UAVs to learn the caching policy, and the contents cached at A-UAVs emulate the cache pre-loading segmentation behavior described in Section IIIA. However, the caching policy and corresponding content availability may fluctuate due to less request for less popular content, leading to weak or unstable reward estimates. This results in Q-values that are highly sensitive to requests for less popular content and less sensitive to requests for popular content. Therefore, changes in Q-values of less popular content may lead to intermittent variations in caching, particularly in Segment-2 (refer Section IIIA). Also, there can be $\binom{N}{k}$ combination of contents to be sampled by the *Top-k* MAB agent for caching. Due to this the reward estimation for each content occurs after large intervals, which leads to a weak estimate of reward distribution as $N$ increases. These oscillations can be controlled by empirically selecting $\epsilon$ and its decay rate. To reduce the dependence of caching policy on the choice of $\epsilon$, Upper Confidence Bound (UCB) strategy is used [1, 11]. The *Top-k* MAB agent maintains an upper confidence bound on the expected reward of each content, and selects the set of $C_A$ contents with highest UCB at each epoch.

$$U_t(i) = Q_t(i) + \sqrt{\frac{\alpha_u \log(t)}{N_t(i)}} \quad (11)$$

Here, $U_t(i)$ is the UCB of content '$i$' at epoch '$t$'; $Q_t(i)$ is the updated Q-value at epoch '$t$'; $\alpha_u$ is a hyperparameter that controls the degree of exploration; $N_t(i)$ is the number of time content '$i$' has been requested till epoch '$t$'. The first term represents the reward estimate, and the second term depicts the uncertainty in reward estimate. UCB selects the content that has high potential for high reward but hasn't been requested frequently. The promotes exploration without externally inducing an exploration parameter such as $\epsilon$. For this paper, $\epsilon$-greedy exploration strategy is applied according to the UCB values, as shown in Step 7-16 in Algorithm 1.

The following pseudo code explains the caching policy at a A-UAV with a *Top-k* MAB agent.

1. **Initialization:**
   a. N: Total contents in the system
   b. $C_A$: Caching capacity of an A-UAV
   c. $Q$: Array of size $C_A$ initialized with 0's (Q-table).
   d. $\epsilon$: Exploration rate
   e. $\alpha$: Learning rate for Q-table update.
   f. $\alpha_u$: Degree of exploration (if UCB used)
2. **Load** A-UAV's cache with $C_A$ randomly chosen contents.
3. **while** True:
   \\ Check for learning epoch
4.   **if** F-UAV is visiting A-UAV **then do**
5.     **for** $i = 0\ to\ length$(A-UAV cache size $C_A$) **do**
6.       **Get** reward $r(i)$ \\ according to Eqns. 7-9
7.       **Update** $Q(i)$ \\ $Q(i) \leftarrow Q(i) + \alpha[r(i) - Q(i)]$
   \\ $Q(i) \leftarrow U(i)$ if UCB employed
8.     **end for**
9.     $value = \boldsymbol{copy}(Q)$ \\ make a copy of Q-table
   \\ Reload contents (Select arms)
10.     **for** $i = 0\ to\ length$(A-UAV cache size $C_A$) **do**
11.       **Generate** random number '$x$'
12.       **if** $x < \epsilon$ **then do**
13.         **Load** 1 randomly chosen content to A-UAV
14.       **else**
15.         $c_{max} = \boldsymbol{argmax}(value)$
16.         **Load** $c_{max}$ to A-UAV
17.         **Set** $value[c_{max}] = -inf$
18.       **end if**
19.     **end for**
20.   **end if**
21.   **Check** for $\epsilon$ decay condition.
22.   **if** true **then do**

```
        Update ϵ
    end if
end while
```

Algorithm 1. Caching policy at a A-UAV with *Top-k* MAB Learning

This *Top-k* MAB agent at a A-UAV learns a near optimal caching policy within a finite time horizon and approaches the best caching policy asymptotically. The cached contents can boost content availability at their respective communities as well as at other distant communities via F-UAVs.

## V. Experiments and Results

Experiments are performed to analyze the performance of the proposed *Top-k* MAB learning-based caching mechanism with a discrete event simulator. The simulator accomplishes content request generation while maintaining an intra-event interval according to exponential distribution and following a Zipf popularity distribution (refer Eqn. 1). To perform the cache pre-loading, the mathematical expressions are included in the simulator. To capture heterogeneity in content popularity sequence at different communities, contents are swapped with pre-decided probability [10] and the difference between the sequences are determined using Smith-Waterman Distance [10]. The experimental parameters for the proposed *Top-k* MAB learning based caching and cache pre-loading policies are listed in Table I. The performance evaluation of the proposed mechanism is accomplished via the following metrics.

TABLE I. Default Values for Model Parameters

| # | Variables | Default Value |
|---|---|---|
| 1 | Total number of contents, $C$ | 1000 |
| 2 | Number of A-UAVs, $N_A$ | 12 |
| 3 | Number of F-UAVs, $N_F$ | 3 |
| 4 | Cache space in A-UAV, $C_A$ | 100 |
| 5 | Cache space in F-UAV, $C_F$ | 100 |
| 6 | Poisson request rate parameter, $\mu$ | 1 request/sec |
| 7 | Hover time of F-UAV, $T_{Hover}$ | 600 seconds |
| 8 | Transition time of F-UAV, $T_{Transition}$ | 300 seconds |
| 9 | Zipf parameter (Popularity), $\alpha$ | 0.7 |
| 10 | Ferrying UAV Trajectory | Round-robin |

*Content Availability* ($P_{avail}$): It is defined as the ratio between cache hits and generated requests within a time interval. Cache hits are the content provided to the users from the contents cached in the UAV-aided caching system (without download). Therefore, content availability indirectly indicates the content download cost of a systems as well.

*Jaro-Winkler Similarity* ($JWS$): It is a similarity measure that is used to compute the similarity between two sequences [12]. It is computed by calculating the number of matches, number of transpositions requires within the matches and the similarity in prefix of both sequences. $JWS$ is used to compute the similarity between the content sequence from the learnt caching policy and content sequence according to cache pre-loading.

*Access Delay* ($AD$): Performance of *Top-K* MAB model is also evaluated based on the access delay which is the end-to-end delay between the generation of content request and its provisioning form the cached contents in the UAVs. This paper reports the epoch-wise average access delay to show the improvement in caching policy as learning progresses.

### A. Effect of Exploration Strategies on Learnt Caching Policy

In order to understand the viability of the proposed *Top-k* MAB learning-based caching policy in scenarios with demand heterogeneity, two type of content popularity sequence are used. Every consecutive community has a different popularity sequence. For $\epsilon$-greedy strategy, initial exploration is $\epsilon = 1$ with decay rate of 0.0025 per epoch. The degree of exploration in UCB is set to $\alpha_u = 2$. Fig. 3a shows the convergence behavior of the learnt caching policy with a comparison of exploration strategies employed in the *Top-k* MAB model.

The convergence behavior is shown in term of content availability from the learnt caching policy. The observations from Fig. 3(a) are as follows. First, the figure shows that by employing *Top-k* MAB agent at every A-UAV, a near optimal caching policy can be learnt. The algorithm is able to leverages the multi-dimensional reward structure, as explained in Eqns. 7-9, to achieve content availability close to the benchmark performance (see section II). Second, when the agent uses UCB exploration strategy, the content availability settles at a sub-optimal value. However, during the initial learning epochs the content availability increases promptly due to high upper confidence value of all contents, which avoids exploitation. This is due to low sampling of requests. As learning progresses, the sparse request for unpopular contents keeps the upper confidence value high which maintains consistent exploratory behavior. An algorithmically induced $\epsilon$ value in $\epsilon$-greedy strategy avoids this continuous uncertainty behavior due to $\epsilon$ decay. This can be seen from the content availability with $\epsilon$-greedy exploration strategy which is better than the performance with UCB. Finally, to maintain the initial surge in content availability and to limit the unbounded exploratory behavior, $\epsilon$-greedy exploration is applied on the UCB values of the content. It can be seen that such hybrid exploration strategy helps to boost the content availability closer to the benchmark performance by 5%. Similarly, Fig. 3(b) shows the convergence behavior of the *Top-k* MAB learning-based caching agent in terms of access delay. This is computed for a $TAD$ of 300 seconds and it is observed that as learning progresses, the access delay for requested contents reduce while the content availability increases simultaneously. This manifests the improvement in learnt caching policy over the learning epochs. The best reduction in access delay is observed when $\epsilon$-greedy exploration is applied on the UCB values of the content.

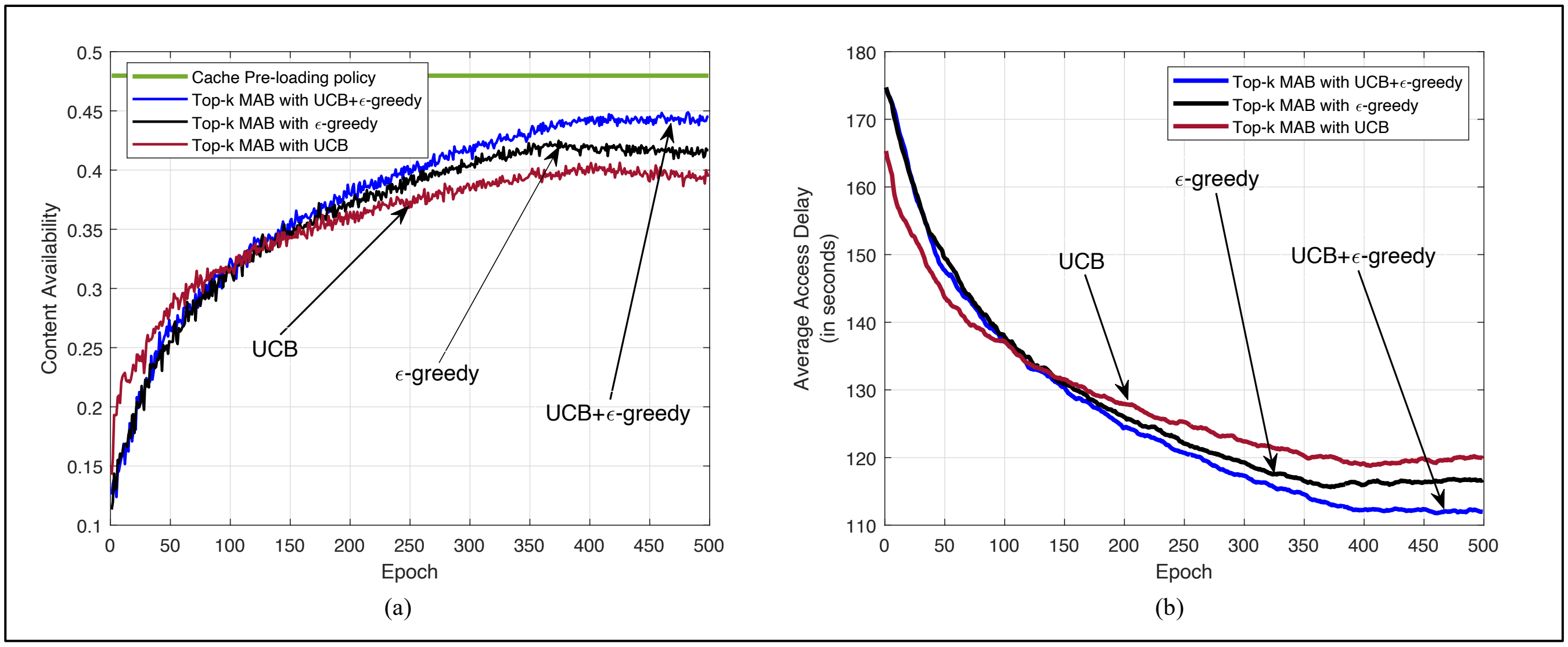

Fig. 3 Comparison between exploration strategies in *Top-k* MAB and pre-loading using (a) Content Availability; (b) Access Delay

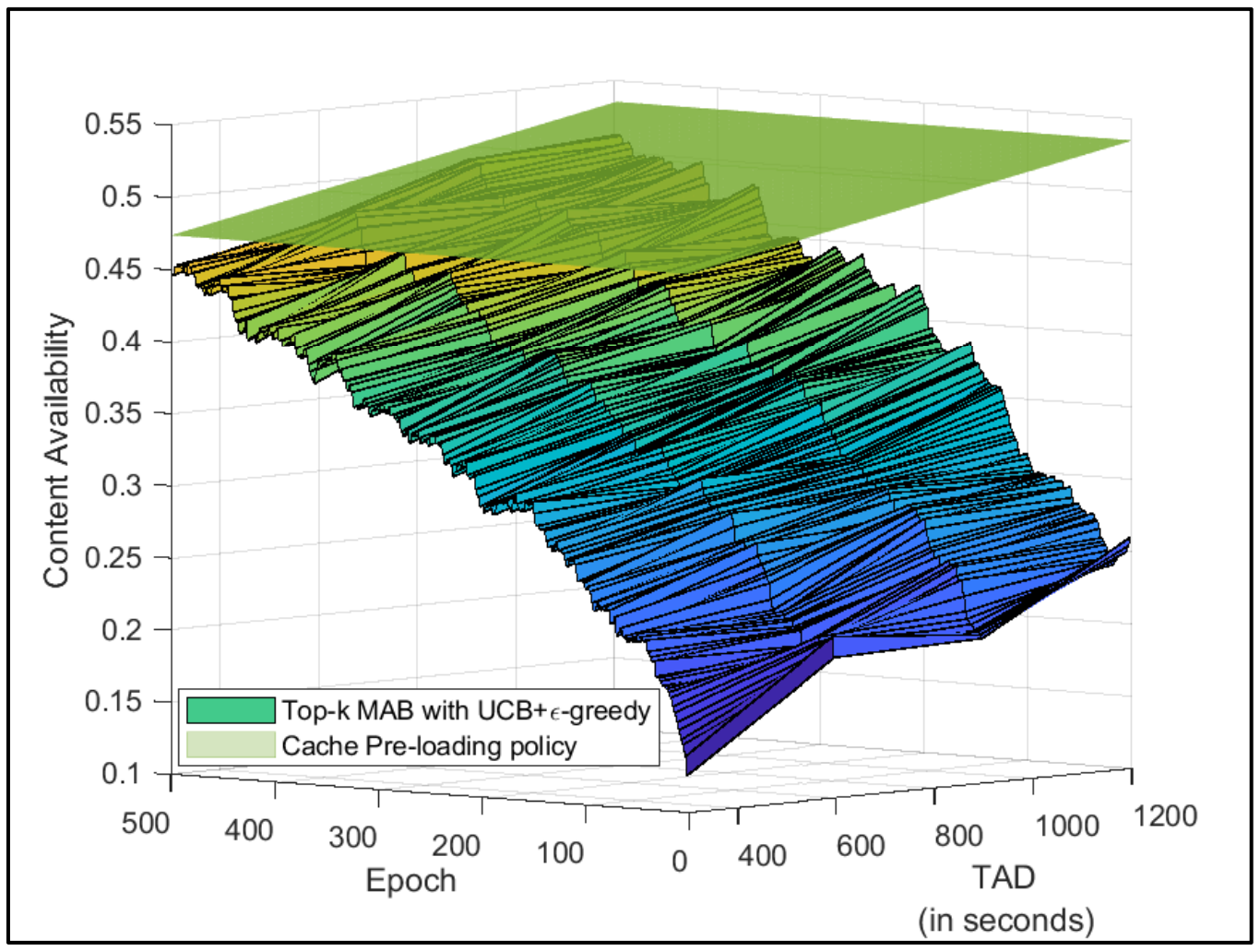

Fig. 4. Change in learnt caching policy of A-UAV with *TAD*

### B. *Impact of Tolerable Access Delay on Learning Performance*

To show the learning capability of the proposed *Top-k* MAB model, experiments are conducted with varying $TAD$s ranging from 300 to 1200 seconds. The content availability according to the learnt caching policy with varying $TAD$ is shown in Fig. 4. The figure demonstrates the behavior of the proposed caching mechanism with respect to the benchmark performance, computed from the cache pre-loading policy discussed in Section II. Following observations can be made from Fig. 4. First, the learnt caching policy achieves performance closer to the benchmark for all values of $TAD$. Second, the best possible performance (i.e., the benchmark) changes with change in $TAD$. The *Top-k* MAB agents in the A-UAVs adapts to the user defined $TAD$. It can be observed in Fig. 4 that the learning performance varies along with $TAD$. In other words, the role of multi-dimensional reward structure of the MAB agent becomes more evident with higher $TAD$. Especially, the information related to the global availability i.e., $\delta_f$ and $\delta_g$ (refer Section IVB), are derived from large count of content requests. This improves the estimated reward at A-UAVs thus impacting their caching decision.

### C. *Cache Similarity of Learnt Sequence with Best Sequence*

The effect of learning on the cached content sequence is demonstrated in Fig. 5. Fig. 5(a) plots Jaro-Winkler Similarity ($JWS$) of cached content sequences for all 12 A-UAVs. The key observation are as follows. First, the $JWS$ between the best caching sequence from cache pre-loading policy (see Section II) and the cached content sequences learnt by the *Top-k* MAB agents at A-UAVs converge near 0.9, with a certain variance. Physically, this represents higher degree of similarity post convergence, where 1 indicates complete similarity and 0 implies no similarity. Second, the cached contents improve over epochs as learning progresses. Lower $JWS$ values at initial epochs signifies that A-UAVs have no *a priori* content popularity information, local or global. As the MAB agents learn, over epochs of generated content requests, the cached contents in A-UAVs become more similar to the best caching sequence. Third, $JWS$ is an indirect representation of the storage segmentation factor ($\lambda$), which is used to decide the segment sizes according to cache pre-loading policies. A higher $JWS$ implies that, along with learning the caching policy, the *Top-k* MAB agents learn to emulate the said segmentation behavior. Finally, the partial dissimilarity of the cached content sequence can be ascribed to the uncertainty associated with the Q-values of contents with low popularity. Also, this leads to an oscillatory convergence of $JWS$ for A-UAVs. This behavior manifests in the $JWS$ for F-UAVs as well, which is shown in Fig. 5(b). Since, F-UAVs ferry contents that are requested less frequently, the low popularity of such contents leads to a comparatively sluggish improvement of its $JWS$ as compared to $JWS$ improvement of A-UAVs.

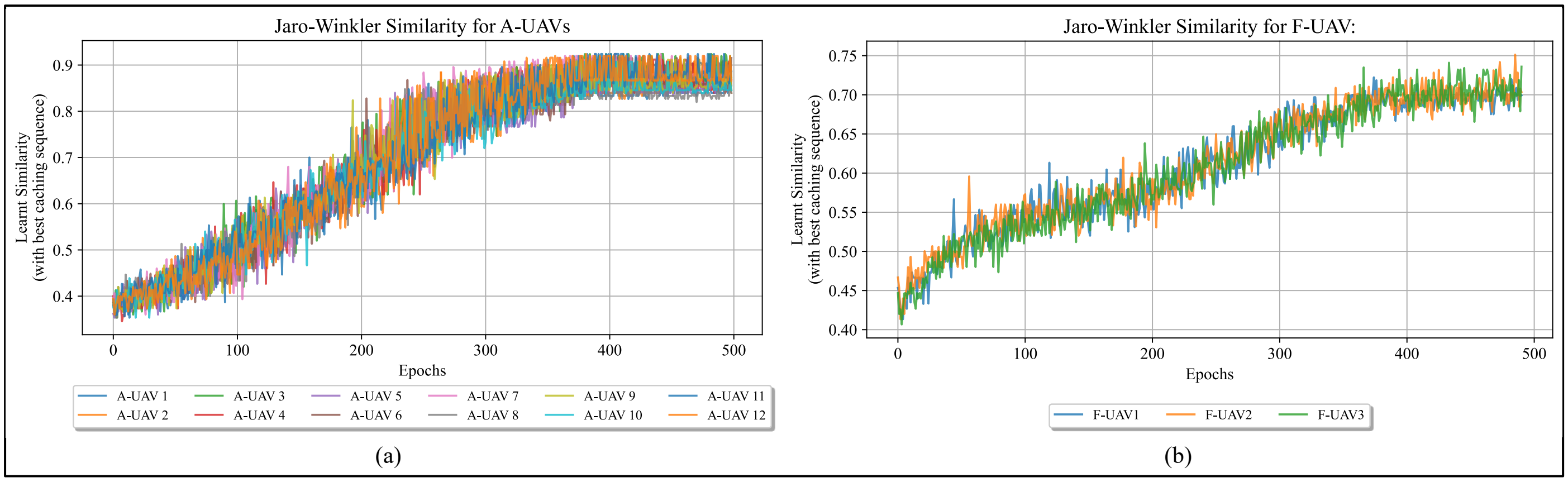

Fig. 5. Jaro-Winkler similarity for A-UAVs and F-UAVs

## VI. Related Work

In recent years, a significant amount of research has been conducted on UAV caching. Such works can be broadly classified into two categories, namely, platform enhancements and algorithmic optimization. UAV platform-related enhancements, however, are not in line with the objectives of this specific paper, which approaches the cache optimization problem in an algorithm-centric manner.

From an algorithmic perspective, authors in [3] show that the effective caching capacity of UAVs can be enhanced by considering the popularity and size of the content being stored. The study in [4] proposes a UAV-enabled small-cell network in which data traffic is offloaded from the of small-cell base-stations (SBSs) to UAVs. The most popular contents are proactively cached within the UAVs, and delivered to the user directly as needed. [5] did similar work where they attempt to reduce the traffic load on ground base-stations via UAV-caching. The approach in [6] uses a joint caching and UAV trajectory optimization using particle swarm optimization by modeling each caching strategy as a particle. In a similar joint optimization study, [7] uses a reinforcement learning-based approach for UAV-caching decision-making where the content requests, storage, and availability in the storage buffer are used for defining states in a Markov Decision Process.

While some of these UAV-based caching mechanisms [4, 5] are useful for partial infrastructure destruction, they are less likely to work well when all communication infrastructures are destroyed, and a fully functional alternative is needed. Additionally, most of the above mechanisms [3-5] consider temporally static global content popularity [5], which misses capture the real-world heterogeneity and time-variability of content demands in disaster scenarios. The optimization mechanisms in [6, 7] use long-term estimation methods which fundamentally lack the promptness and adaptability with changing network and demand conditions. Explicit attempts for effective cache space maximization, and reduction of expensive from-server downloads using vertical links are also absent in the prior published works on UAV caching.

To addresses those issues, the *Top-k Multi-Armed Bandit* learning model is developed for UAV-caching decisions that take geo-temporal differences in content popularity and heterogeneity in demand into consideration. The approach employs a multi-dimensional reward structure that improve system performance via sharing information between UAVs.

## VII. Summary and Conclusion

In this paper, UAV-aided content dissemination system is proposed which can learn the caching policy on-the-fly without *a priori* content popularity information. Two types of UAVs are introduced to revive content provisioning in a disaster/war-stricken scenario viz. anchor and ferrying UAVs. Cache-enabled anchor UAVs are stationed at each stranded community of users for uninterrupted content provisioning. Ferrying UAVs act as content transfer agents across anchor UAVs. The evolution of pre-loading-based caching policies are discussed which requires *a priori* information about content popularity. A decentralized *Top-k* Multi-Armed Bandit Learning-based caching policy is proposed to ameliorate the limitation of cache pre-loading. It learns the caching policy on-the-fly with the help of a multi-dimensional reward structure with encapsulates local and global availability information. Future work on this research will include distributed learning model sharing approaches to improve content provisioning.

## VIII. References

[1] Slivkins, A. (2019). Introduction to multi-armed bandits. *Foundations and Trends® in Machine Learning*, *12*(1-2), 1-286.

[2] Zeng, Y., et al. (2016). Wireless communications with unmanned aerial vehicles Opportunities and challenges *IEEE Communications magazine 54*(5) 36-42.

[3] Mozaffari, M., et al. (2019). A tutorial on UAVs for wireless networks: Applications, challenges, and open problems. *IEEE communications surveys & tutorials*, *21*(3), 2334-2360.

[4] Zhao, N., et al. (2019). Caching unmanned aerial vehicle-enabled small-cell networks: Employing energy-efficient methods that store and retrieve popular content. *IEEE Vehicular Technology Magazine*, *14*(1), 71-79.

[5] Zhang, T., et al. (2020). Cache-enabling UAV communications: Network deployment and resource allocation. *IEEE Transactions on Wireless Communications*, *19*(11), 7470-7483.

[6] Wu, H., et al. (2020). Optimal UAV caching and trajectory in aerial-assisted vehicular networks: A learning-based approach. *IEEE Journal on Selected Areas in Communications*, *38*(12), 2783-2797.

[7] Al-Hilo, et al. (2020). UAV-assisted content delivery in intelligent transportation systems-joint trajectory planning and cache management. *IEEE Transactions on Intelligent Transportation Systems*, *22*(8), 5155-5167.

[8] Bhuyan, A. K., et al. (2022, December). Towards a UAV-centric Content Caching Architecture for Communication-challenged Environments. In *2022 IEEE Global Communications Conference* (pp. 468-473).

[9] Bhuyan, A. K., et al. (2023, January). UAV Trajectory Planning For Improved Content Availability in Infrastructure-less Wireless Networks. In *2023 International Conference on Information Networking (ICOIN)* (pp. 376-381).

[10] Bhuyan, A. K., Dutta, H., & Biswas, S. (in press). Handling Demand Heterogenity in UAV-aided Content Caching in Communication-challenged Environments. In *2023 IEEE 24th International Symposium on "A World of Wireless, Mobile and Multimedia Networks" (WoWMoM)* (pp. xxx-xxx). IEEE.

[11] Cao, W., et al. (2015). On top-k selection in multi-armed bandits and hidden bipartite graphs. *Advances in Neural Information Processing Systems*, *28*.

[12] Cohen, W., et al. (2003, August). A comparison of string metrics for matching names and records. In *Kdd workshop on data cleaning and object consolidation* (Vol. 3, pp. 73-78).